\documentclass[prb,aps,twocolumn,showpacs]{revtex4}
\usepackage{graphicx}
\usepackage{dcolumn}
\usepackage{bm}
\usepackage{color}
\begin{document}

\title{Evolutions of helical edge states in disordered HgTe/CdTe quantum
wells}
\author{Liang Chen,$^{1,2}$ Qin Liu,$^{1}$ \footnote{E-mail: liuqin@mail.sim.ac.cn}
Xulin Lin$^{1,2}$, Xiaogang Zhang$^{1,2}$, Xunya Jiang$^{1}$ }
\affiliation{ $^1$State Key Laboratory of
Functional Materials for Informatics, Shanghai Institute of
Microsystem and Information Technology, CAS, Shanghai 200050, China}
\affiliation{ $^2$Graduate School of Chinese Academy of Sciences,
Beijing 100049, People's Republic of China}
\date{\today}

\begin{abstract}
We study the evolutions of the nonmagnetic disorder-induced edge
states with the disorder strength in the HgTe/CdTe quantum wells.
From the supercell band structures and wave-functions, it is clearly
shown that the conducting helical edge states, which are responsible
for the reported quantized conductance plateau, appear above a
critical disorder strength after a gap-closing phase transition.
These edge states are then found to decline with the increase of
disorder strength in a stepwise pattern due to the finite-width
effect, where the opposite edges couple with each other through the
localized states in the bulk. This is in sharp contrast with the
localization of the edge states themselves if magnetic disorders are
doped which breaks the time-reversal symmetry. The size-independent
boundary of the topological phase is obtained by scaling analysis,
and an Anderson transition to an Anderson insulator at even stronger
disorder is identified, in-between of which, a metallic phase is
found to separate the two topologically distinct phases.
\end{abstract}
\pacs{73.43.Nq, 72.25.-b, 71.30.th, 72.15.Rn}

\maketitle

\section{ Introduction}
Disorder effect is one of the most important problems in the subject
of topological insulators (TI) \cite{Review}. Recently a nonmagnetic
{\it disorder-induced} topological insulator state, named as the
topological Anderson insulator (TAI), is found by computer
simulations in both two- \cite{Li2009PRL} and three-dimensional (3D)
\cite{Guo2010PRL} systems, and is further confirmed through
independent simulations by studying the local currents
\cite{Jiang2009PRB}. This state is numerically characterized by
precisely quantized conductance $e^2/h$
\cite{Li2009PRL,Guo2010PRL,Jiang2009PRB}, and is theoretically
understood as generated by the negative renormalized topological
mass \cite{Groth2009PRL,Guo2010PRL} within the self-consistent Born
approximation (SCBA). Phase diagrams for these systems on the
energy-disorder plane at fixed mass parameter are obtained
\cite{Li2009PRL,Guo2010PRL,Groth2009PRL}, where the TAI phase
appears as an island between the so-called weak- and strong-disorder
boundaries \cite{Groth2009PRL}. The locations of these boundaries
depend generally on the size of the system under simulation.
Topological invariants in disordered systems are also discussed
\cite{Prodan2011,Guo2010PRB}. In particular, it is argued that the
2D TAI phase is not a distinct one, but is instead part of the
quantum spin Hall phase \cite{Prodan2011} with nontrivial spin-Chern
number \cite{Prodan2011B} in the presence of disorder if the phase
diagram is extended to the mass axis.

Despite all these achievements on understanding the surprising
disorder-induced topological phase, there still exist some
questions. First, it has been experimentally proved that the
quantized conductance, $2e^2/h$, in clean 2D HgTe/CdTe quantum wells
(QWs) \cite{Bernevig2006} measured in transport studies
\cite{Konig2007} is the direct consequence of gapless conducting
edge states. However in the disorder-induced TI, the existence of
such edge states in energy-momentum space, their evolution with
disorder strength, as well as how the evolution picture corresponds
to the wave-function behaviors in real space have not been well
investigated yet. Second, when deviating the TAI region, how these
edge states terminate with the increase of the disorder strength and
what is their difference from that of the usual bulk states are also
not resolved so far. Finally, as mentioned in
Ref.\cite{Groth2009PRL}, the weak-disorder boundary is not an
Anderson transition at all as the name TAI might suggest, in the
sense that it is more of a band-effect rather than the result of a
mobility edge. Therefore the Anderson transition boundary, which is
expected to be the true size-independent phase boundary of TAI in
strong-disorder region, is still absent for now.

\begin{figure}[tbp]
\includegraphics[width=1.0 \columnwidth]{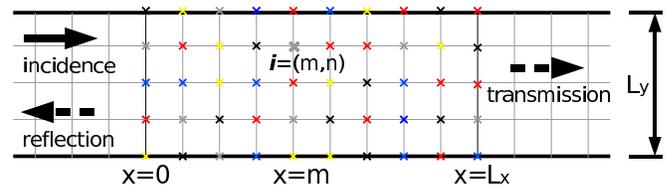}
\caption{(color online). Schematic illustration of the two-terminal
device. The central stripe is made of HgTe/CdTe QWs with length
$L_x$ and width $L_y$ where nonmagnetic disorders exist. Two
semi-infinite clean leads made of the same materials are fabricated
which are heavily doped to access the $|E_f|<M$ region
\cite{Groth2009PRL}.} \label{Structure}
\end{figure}
Motivated by these issues, in this paper, we study the
disorder-induced conducting edge states and the phase boundaries of
the TAI in the effective model of disordered HgTe/CdTe QWs. First,
the existence of gapless helical edge states is demonstrated in
energy-momentum space through a gap-closing phase transition when
above a critical disorder strength. This result appraises the
topological aspect of the TAI in a straightforward way, and is the
fundamental reason of the observed quantized anomalous conductance
plateaus in previous transport studies
\cite{Li2009PRL,Jiang2009PRB,Guo2010PRL,Groth2009PRL}. Second, the
evolution of such disorder-induced edge states with disorder
strength is also studied, where how the edge states decay in
strong-disorder region is focused on in particular. It is shown that
for a finite-width system, unlike the usual exponential decay in
quasi-1D disordered systems \cite{MacKinnon}, the nontrivial edge
states decay in a {\it stepwise} pattern due to the competition
between the sample width $L_y$ and the localization length $\xi(W)$.
Finally, we locate the true Anderson transition boundaries by
scaling analysis for an infinite system. These boundaries indicate
that the system undergoes a multiple transitions, first from metal
to topological insulator, then to metal again, and finally localizes
as an Anderson insulator. It is interesting to compare with the work
by Yamakage \cite{Yamakage2010} {\it et al} that our result shows
that a spin-$s_z$ nonconservation term is not necessarily needed to
have a metallic region which separates two topologically distinct
insulating phases.

The rest of this paper is organized as follows. In section II, we
introduce our simulating model and methods. In section III, we
present our numerical results and theoretical analyses. Finally,
this work is concluded in section IV.

\section{ Model and method}
Our starting point is the 2D effective Hamiltonian of HgTe/CdTe QWs
\cite{Bernevig2006} with on-site nonmagnetic disorders. In
tight-binding representation on a square lattice, it gives
\cite{Jiang2009PRB}
\begin{eqnarray}
H=\sum_{\bf i}\epsilon_{\bf i}\psi_{\bf i}^{\dagger}\psi_{\bf i}+
(t_x \psi_{\bf i}^{\dagger}\psi_{{\bf i},{\bf i}+\hat x}+ t_y
\psi_{\bf i}^{\dagger}\psi_{{\bf i},{\bf i}+\hat y}+h.c.),\label{TB}
\end{eqnarray}
where $\psi_{\bf i}=(\psi_{{\bf i}s\uparrow}, \psi_{{\bf
i}p\uparrow}, \psi_{{\bf i}s\downarrow}, \psi_{{\bf
i}p\downarrow})^{\rm T}$ is the field for the four orbital states,
$\left|s,\uparrow\right>$, $\left|p_x+ip_y,\uparrow\right>$,
$\left|s,\downarrow\right>$, $\left|-p_x+ip_y,\downarrow\right>$, on
site ${\bf i}=(m,n)$, and $\epsilon_{\bf i}$ and $t_{x(y)}$ are
respectively the on-site energy as well as the overlap integral
matrices along $x(y)$ direction, which are explicitly
\begin{eqnarray}
&&\epsilon_{\bf i}=\rm Diag( \epsilon_{s}+\Delta_{\bf
i},\epsilon_{p}+\Delta_{\bf i},\epsilon_{s}+\Delta_{\bf
i},\epsilon_{p}+\Delta_{\bf i}),\\
&&t_x= \left(
\begin{array}{cccc}
 \frac{D+B}{a^2} & \frac{-iA}{2a} & 0 & 0\\
 \frac{-iA}{2a} & \frac{D-B}{a^2} & 0 & 0\\
 0 & 0 & \frac{D+B}{a^2} & \frac{iA}{2a}\\
 0 & 0 & \frac{iA}{2a} & \frac{D-B}{a^2}
\end{array}
\right),\\
&&t_y= \left(
\begin{array}{cccc}
 \frac{D+B}{a^2} & \frac{A}{2a} & 0 & 0\\
 \frac{-A}{2a} & \frac{D-B}{a^2} & 0 & 0\\
 0 & 0 & \frac{D+B}{a^2} & \frac{A}{2a}\\
 0 & 0 & \frac{-A}{2a} & \frac{D-B}{a^2}
\end{array}
\right).
\end{eqnarray}
In the above, $a$ is the lattice constant, $A$,$B$,$C$,$D$,$M$ are
material parameters depending on the QW width \cite{Bernevig2006},
$\epsilon_{s(p)}=C\pm M-4(D\pm B)/a^2$, and $\Delta_{\bf i}$ is the
on-site disorder energy for nonmagnetic impurities, which is
identical for the four orbitals and uniformly distributed in
$[-W/2,W/2]$ with disorder strength $W$. To compare with the
previous works, we use the same values for the parameters as in
Refs. \cite{Li2009PRL,Jiang2009PRB}, i.e., $A=364.5$ meV$\cdot$nm,
$B=-686$ meV$\cdot$nm$^2$, $C=0$, $D=-512$ meV$\cdot$nm$^2$, $M=1$
meV and $a=5$ nm.
\begin{figure*}[tbp]
\includegraphics[width=18cm]{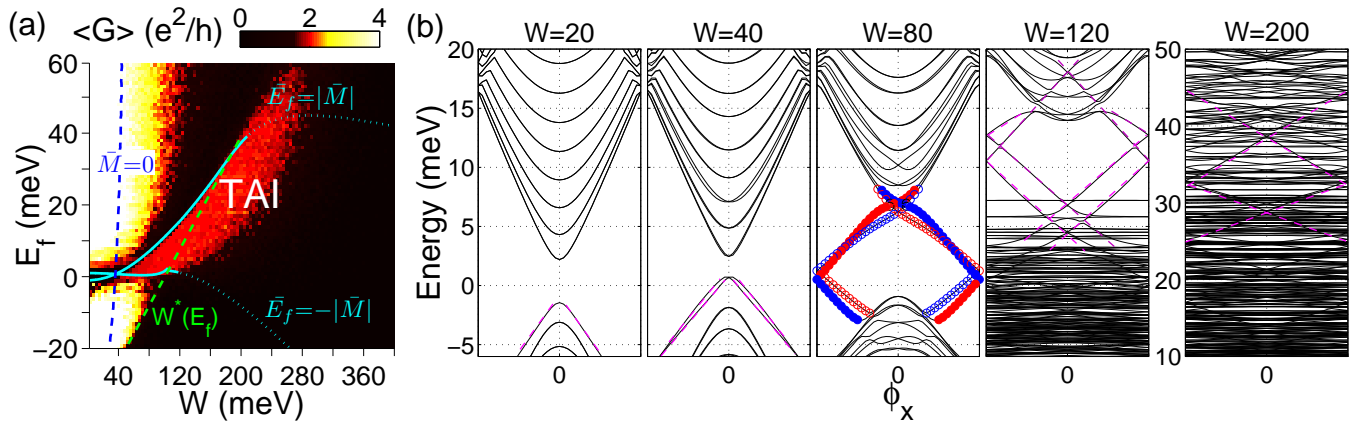}
\caption{(color online). (a) Phase diagram of disordered HgTe/CdTe
QWs on $E_f-W$ plane, which is simulated from a stripe geometry with
$L_x=500a$ and $L_y=100a$. The curves are theoretical results
obtained from the SCBA. The blue dashed line represents the
interface $\bar{M}=0$, to the right of which the mass parameter is
renormalized to negative. The green dashed line is the border where
the self-energy becomes imaginary. The TAI phase is indicated as the
crescent-shaped region where the phase boundaries,
$|\bar{E_f}|=\pm|\bar{M}|$, are marked by the solid cyan lines. It
is seen that the SCBA fails at strong-disorder boundary as denoted
by the dotted parts. (b) The evolutions of TAI's energy spectrum with
disorder strength $W$ in the generalized momentum space. In the
subfigure of $W=80$, the edge states are labeled according to their
density distributions (empty and solid circles) as well as the spin
polarization (blue and red). The purple dashed lines in the other
subfigures are guides for the eyes of the edge states dispersions.}
\label{EnergySpectrum}
\end{figure*}

The method of twisted boundary conditions \cite{Niu1985,Qi2006} is
used in this work to directly diagonalize the disordered system, and
the transfer matrix method \cite{MacKinnon} is taken to obtain the
wave-function distributions.

Without disorder, the mass parameter $M$ characterizes the effect of
band inversion by changing its sign from positive to negative, and
results in a topological phase transition from a conventional
insulating state ($M>0$) to a quantum spin Hall phase ($M<0$) with a
single pair of helical edge states at each boundary. This pair of
helical edge states is clearly seen in energy-momentum space by
diagonalizing the clean Hamiltonian (setting $\Delta_{\bf i}=0$ in
Eq.(\ref{TB})). However, for disordered systems, the translational
symmetry is generally broken and the momentum is not a good quantum
number anymore. But if we view the disordered system of size
$L_x\times L_y$ as a cell of an infinite one, and fold it into a
cylinder with longitudinal axis in $y$-direction, then the
translational symmetry is restored, where a particle gains an extra
phase factor $e^{i\phi_x}$ whenever it goes across the boundary in
the $x$-direction. In such a case, $-\pi\leq\phi_x\leq\pi$, plays
the role of the generalized momentum, and we could diagonalize the
disordered system in $E-\phi_x$ plane and study the edge properties
as a function of $W$. It should be noted that the choice of $L_x$
should be larger than the average decay length of the
wave-functions. It is found that $L_x=15a$ is good enough for
diagonalization, and we set $L_y=100a$ all throughout the paper
unless mentioned in particular.

Wave-functions study takes its special advantages to gain insights
of how a conventional insulating state transits into a topological
nontrivial phase and vice versa. To obtain the wave-functions in
real space, we consider a two-terminal setup as seen in
Fig.\ref{Structure}, and adopt the standard transfer matrix method
\cite{MacKinnon} with the iteration equation along $x$ direction
written as
\begin{eqnarray}
\left(
\begin{array}{c}
\Psi(m+1)\\
\Psi(m)
\end{array}
\right) = M(m) \left(
\begin{array}{c}
\Psi(m)\\
\Psi(m-1)
\end{array}
\right),
\end{eqnarray}
where $\Psi(m)=(\psi_{m1}, \psi_{m2},\cdots, \psi_{mL_y})^{\rm T}$
is the wave-function vector at slice $x=m$, and $M(m)$ is the
corresponding transfer matrix. In the region of disordered HgTe/CdTe
QWs, the transfer matrix reads explicitly
\begin{eqnarray}
M(m)= \left(
\begin{array}{cc}
{T_x}^{-1}[E_f I-H(m)] & -{T_x}^{-1} {T_x}^{\dagger}\\
I & O\\
\end{array}
\right),
\end{eqnarray}
where $E_f$ is the Fermi energy of the clean system, $I$ and $O$ are
respectively the $4L_y\times 4L_y$ unit and null matrices, and
\begin{eqnarray}
T_x=\rm{Diag}(t_x, t_x,\cdots,t_x)
\end{eqnarray}
is the block diagonal overlap matrix with the same dimension, while
\begin{eqnarray}
H(m)= \left(
\begin{array}{cccccc}
\epsilon(m,1) & t_y &  &  &  & \\
t_y^{\dagger} & \epsilon(m,2) & t_y &  &  & \\
 & . & . & . &  & \\
 &   & t_y^{\dagger} & \epsilon(m,n) & t_y & \\
 &   &   & . & .  & .\\
\end{array}
\right)
\end{eqnarray}
is the block-tridiagonal slice Hamiltonian with open boundary
conditions in $y$-direction. By applying a small external bias, we
can calculate the longitudinal conductance using the
Landauer-B\"uttiker formulae \cite{MacKinnon,Landauer1970} as well
as the wave-functions on each site. Fig.\ref{EnergySpectrum}(a)
shows the TAI phase diagram reproduced by the transfer matrix
method, which is in good consistent with the previous works
\cite{Li2009PRL,Groth2009PRL}.

In the following, we first explore the evolution of the energy
spectrum in the generalized momentum space with disorder strength
$W$, and then study the topological phase transition in the
viewpoint of real space wave-functions. The obtained results agree
with each other well, which also validates our numerical methods.

\section {Numerical Results}
\begin{figure*}[tbp]
\includegraphics[width=18cm]{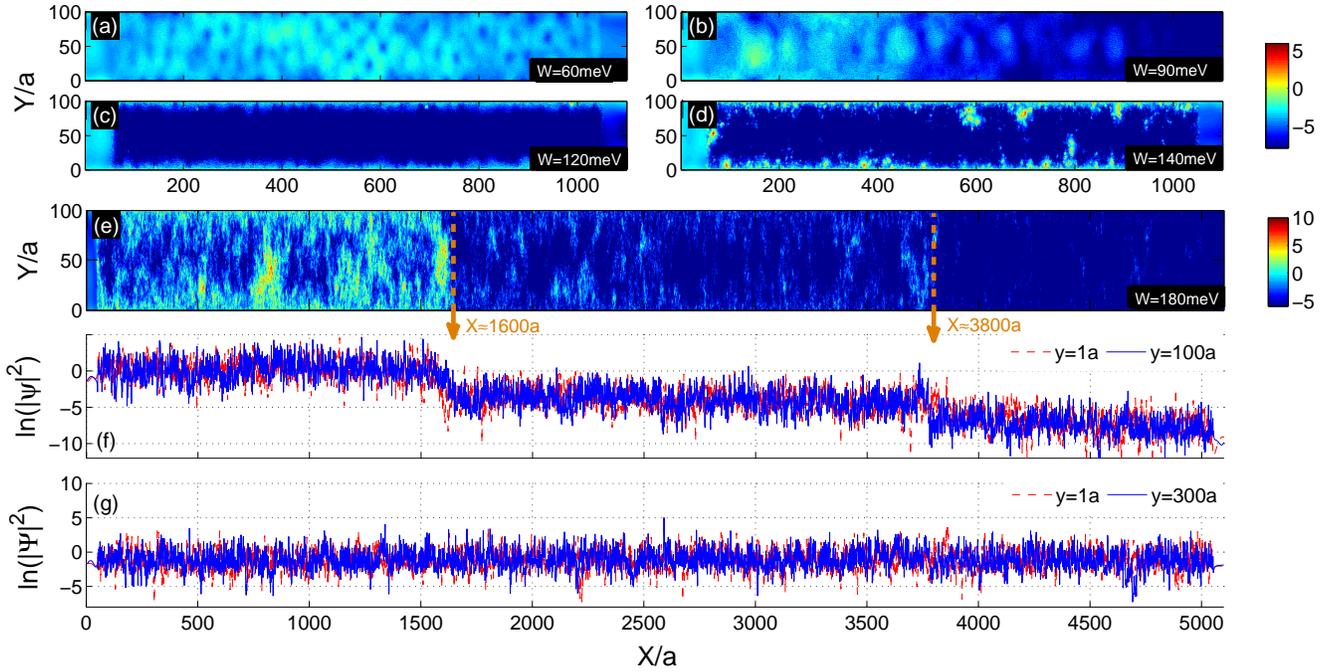}
\caption{(Color online). Logarithmic wave-function intensities at
$E_f=10$ meV. The wave-function intensities before, on, and at the
tail of the TAI plateau are shown respectively in (a)-(b), (c)-(d),
and (e). The edge states wave-function intensities in (e) at $y=1a$
and $100a$ are given in (f), where the brown arrows indicate the
positions where accidental couplings between the edge states occur
due to the large fluctuations of the local decay length. For comparison,
the edge states wave-function intensities of a wider stripe with
$L_y=300a$ are plotted in (g) with all the other parameters remain
the same as in (e). In all the figures, the first and last 50
lattices are leads region.} \label{WaveFunction}
\end{figure*}

\subsection{Overview of the phase diagram}
Let us first review the numerical and theoretical results of the TAI
phase diagram \cite{Li2009PRL,Groth2009PRL}, which is reproduced in
Fig.\ref{EnergySpectrum}(a) with the stripe length $L_x=500a$.

Numerically the TAI phase is identified by the suddenly appeared
quantized conductance, $2e^2/h$, as the increasing of $W$ at fixed
$E_f$. See the crescent-shaped region in
Fig.\ref{EnergySpectrum}(a), where the statistical fluctuations are
vanishing small. This region has a clear-cut weak-disorder boundary
at which the TAI phase begins but a blurred strong-disorder boundary
where the TAI phase terminates.

Theoretically the transitions into the TAI phase at the
weak-disorder boundary can be understood within the the SCBA
\cite{Groth2009PRL,Guo2010PRL}. It is demonstrated that the random
potential renormalizes both the Fermi energy and the mass parameter
as a function of disorder strength, $\bar{E}_f(W)$ and $\bar{M}(W)$,
so that starting from a metallic phase, the mass changes sign when
acrosses some critical disorder strength, and the TAI phase is
defined by the combination of conditions $\bar{M}<0$ as well as
$-|\bar{M}|<\bar{E_f}<|\bar{M}|$. In Fig.\ref{EnergySpectrum}(a),
the boundary $\bar{M}(W)=0$, which separates the positive and
negative effective mass is plotted as blue dashed line, and the band
edges, $\bar{E}_f=\pm\bar{M}$, obtained from the SCBA are shown in
cyan. It is seen that the results predicted from the SCBA agree very
well with the weak-disorder boundary of the TAI phase obtained from
our numerical simulations. However the SCBA fails to give the
correct strong-disorder boundaries, as seen by the dotted parts of
the cyan lines.

Physically the strong-disorder boundary is blurred because in this
region there are impurity states enter into the renormalized bulk
gap, which embodies in the property that the self-energy has a
nonvanishing imaginary part \cite{Guo2010PRL}. In
Fig.\ref{EnergySpectrum}(a), the critical disorder strength is
shown as a function of Fermi energy, $W^{\ast}(E_f)$, see the
green dashed line, where for $W>W^{\ast}$ the self-energy becomes
imaginary. This line distinguishes the TAI phase into two regions by
whether the band edges are effectively defined or not. If they are
not, the mobility edges should play the role.

Finally we note that the strong-disorder boundary is expected to be
an Anderson transition boundary
\cite{Groth2009PRL,Guo2010PRL,Guo2010PRB}, however a direct scaling
analysis of Anderson transitions is still absent, which we will
present below.

\subsection{Evolutions of edge states with disorder strength}
As mentioned above, the observed quantized conductance plateau in
TAI phase is attributed to the presence of conducting edge states
induced by disorder. However, the existence of such edge states has
never been shown directly in the TAI's energy spectrum. In this
section, we first diagonalize the HgTe/CdTe QWs for a specific
disorder realization with increasing disorder strength $W$ (meV).
The twisted and open boundary conditions are used respectively in
$x$ and $y$ directions. The obtained energy spectrum is in
$E-\phi_x$ plane, and five representatives with $W=0$, 40, 80, 120
and 200 are given in Fig.\ref{EnergySpectrum}(b)\cite{online}.

In the clean limit where $W=0$, we see that the spectrum exhibits a
topological-trivial feature with a full bulk gap
$E_g\simeq2\left|M\right|$ (the derivations are due to the finite
size of $L_y$), which inversely proves the validity of our method.
With the increasing of the disorder strength, say to $W=40$, it is
seen that though the system still behaves like a normal insulator,
there are trivial edge states (see the dashed purple lines) grown up
from the lower band and the bulk gap becomes much narrower. In fact,
we have observed the closing of band gap around $W_c=50$ (for our
particular disorder realization) and its reopening immediately at a
bit larger disorder strength, but with the presence of two pairs of
gapless edge states. Since the phase transition necessarily
accompanies with gap-closing \cite{Murakami2007}, we speculate that
the system transits into the TAI phase for $W>W_c$, and a typical
spectrum after this transition is exemplified at $W=80$. The
observation of the gapless edge states with nonvanishing disorders
in generalized momentum space provides the most direct evidence so
far that it is the disorder-induced edge transport which leads to
the quantized conductance in TAI phase. Moreover, these gapless edge
states persist to even stronger disorder strength, nevertheless the
low energy states begin to localize and smear the band edges, see
the $W=120$ figure for example, which is in good consistent with the
phase diagram. Interesting thing happens in the strong-disorder
region. With strong-disorder, naively we would expect that all
states are localized and the system becomes an Anderson insulator.
However, the energy spectrum at $W=200$ shows that although the bulk
gap is completely smeared by the impurity states and the band edges
are ill-defined, the edge states are still robust and winding around
the projected Brillouin zone.

To compare with the helical properties \cite{Wu2006} of the edge
states in clean HgTe/CdTe QWs, we have studied the edge states with
disorder in detail by analyzing their eigen-functions. The results
are shown in the sub-figure with $W=80$. The edge states are labeled
according to their density distributions $\rho(y)=\int
dx\Psi^{\dagger}(\vec{i})\Psi(\vec{i})$ as well as the spin
polarization $\vec{s}(y)=\int
dx\Psi^{\dagger}(\vec{i})\vec{\sigma}\Psi(\vec{i})$, where the solid
(empty) circles indicate the states exponentially localized near the
$y=0\;(L_y)$  boundary, and the color red (blue) represents the up
(down) spin polarization. The helical character of these edge states
for a given energy is clearly seen, where there is a single Kramer's
pair at each edge with spin-momentum locking. The whole spectrum
preserves the time-reversal symmetry by observing that
$E_{n\alpha}(\phi_x)=E_{n\bar{\alpha}}(-\phi_x)$, where $n$ is the
band index and $\alpha=\uparrow,\downarrow$, which is a result of
nonmagnetic disorders. Whereas the spin degeneracy is lifted due to
the breaking of inversion symmetry by disorder. At the time-reversal
invariant points, $\phi_x=0,\pi$, the edge states switch their
Kramer's partners \cite{Fu2007PRB}, which leads to a nontrivial
topology. This is also confirmed by calculating the $Z_2$ index
$\nu_0$ of this 2D system with the twisted boundary conditions on
both directions following the method in Ref. \cite{Guo2010PRB}. All
these results strongly support the statement that the transition
from a normal insulator at $W=0$ to TAI is a topological one, and
the quantized conductance plateau originates from the
disorder-induced conducting helical edge states which are intrinsic
to TAI.

\subsection{Edge states destruction at strong disorder}
The robustness of the disorder-induced edge states shown in
Fig.\ref{EnergySpectrum}(b) (the sub-figure with $W=200$) naturally
raises the questions that how will these edge states be destroyed in
strong-disorder region, and what is their difference with the
exponential localization of bulk states? Not like the case where the
edge states themselves could be localized by magnetic perturbations,
with nonmagnetic disorders, the edge states decay only through the
coupling with each other. For a finite-width system, this generally
depends on the competition between the localization length $\xi(W)$
of bulk states and the system width $L_y$. Only if $\xi(W)\sim L_y$,
even localized bulk states can couple the edge states at opposite
boundaries and destroy the edge states, thus we may expect something
interesting to happen. While for an infinite system, only extended
states can couple the edge states far away from each other,
therefore a localization-delocalization Anderson transition is
expected, which is the true strong-disorder boundary of TAI phase.
In this section, we investigate the decay mechanism of the
disorder-induced edge states for a finite-width system by studying
its wave-functions in real space. The scaling analysis of Anderson
transitions for an infinite system is discussed in the next section.
\begin{figure}[tbp]
\includegraphics[width=9cm]{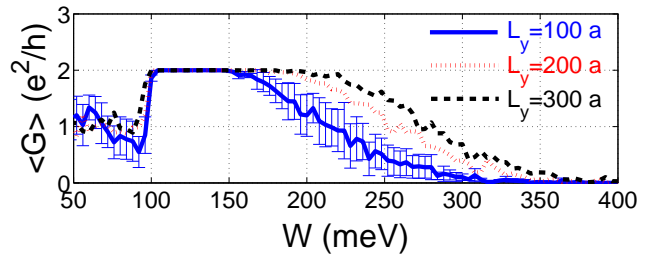}
\caption{\label{fig:epsart}(color online). The average conductance
$\left<G\right>$ versus disorder strength $W$ for different system
width at $E_f=10$ meV. Each line is averaged from $100$ disorder
realizations and $L_x=500a$. The fluctuations are only shown for the
line of $L_y=100a$.} \label{GW}
\end{figure}

We first let $L_x=10^3a$ and fix $E_f=10$ meV. The logarithmic
wave-function distributions before, on, and at the tail of
conductance plateau are given respectively in
Figs.\ref{WaveFunction}(a)-(b), (c)-(d), and (e). Before the TAI
plateau, we see that the bulk states are weakened with the
enhancement of disorder strength. When on the TAI
plateau, the bulk states disappear completely and two
propagating 1D edge states are seen clearly, which contribute a
quantized conductance $e^2/h$ per edge. However, with the increase
of disorder strength, the bulk states reappear and begin to localize
but coexist with the conducting edge states, as seen in
Fig.\ref{WaveFunction}(d). This is equivalent to the $W=120$ case in
Fig.\ref{EnergySpectrum}(b) in the generalized momentum space where
the band edges are ill-defined but the mobility edge starts to
count. We see that the wave-function behaviors in real space agree
very well with the pictures presented above when studying the energy
spectrum.

Interesting thing happens when we move to the tail of TAI plateau,
as seen in Fig.\ref{WaveFunction}(e) where a longer length
$L_x=5\times 10^3a$ is set. With periodic boundary conditions on
both directions, only bulk states exist, which decay exponentially
in the propagating direction \cite{MacKinnon}. Differently here,
with open boundary conditions in one direction, we see that the edge
states fade in three segments as indicated by the brown arrows, and
decay much slower with localized bulk states in between. To check
the interplay between these localized bulk states and the edge
states, we choose two lines at $n=1$ and 100 and plot the
logarithmic intensities of their wave-functions in
Fig.\ref{WaveFunction}(f). It is striking to find that the edge
states decay in a {\it stepwise} pattern in contrast to the
exponential way. This unusual behavior can be understood by the {\it
accidental coupling} between the edge states at opposite boundaries
assisted by the localized bulk states due to the fluctuations of the
decay length $L_d$. It is estimated that
$\xi\simeq\langle\L_d\rangle\sim 10a$ at this disorder strength,
which is comparable to $L_y$, and the intensities of edge states
wave-functions drop off approximately at the interfaces of each
segment, $m=1600$ and 3800, where the local decay length, $L_d(m)$,
is large. To check this, we enlarge the system width to $L_y=300a$,
and the corresponding edge states wave-function intensities at $n=1$
and 300 are given in Fig.\ref{WaveFunction}(g). In this case, we
have $\xi\simeq\langle\L_d\rangle\ll L_y$, therefore a longer system
length is needed for the accidental coupling between the edge states
to happen. For the current simulation, it is seen that the edge
states persist throughout the system, and no coupling occurs at the
much larger scale of $5\times 10^3a$.

\subsection{Scaling analysis of Anderson transitions}
The above wave-function analysis shown in
Figs.\ref{WaveFunction}(f)-(g) implies that the range of the TAI
plateau is system-size dependent. For a fixed length $L_x$, the
wider the width of the system is, the broader the conductance
plateau will be. This is indeed true as seen in Fig.\ref{GW} where
the conductance plateaus for $L_y=100a,\;200a$ and 300$a$ at
$L_x=500a$ are presented. What will happen as $L_y$ tends to
infinite? For an infinite system, if the bulk states keep localized,
where the localization length $\xi$ has a finite value, then the
edge transports, so that the conductance plateaus, will not be
destroyed no matter how large the $\xi$ is. Thus the TAI phase
disappears in an infinite system only when the bulk states become
delocalized and $\xi$ is divergent. Therefore a
localization-delocalization Anderson transition is expected, which
kills the TAI phase in strong-disorder region.

\begin{figure}[tbp]
\includegraphics[width=8cm]{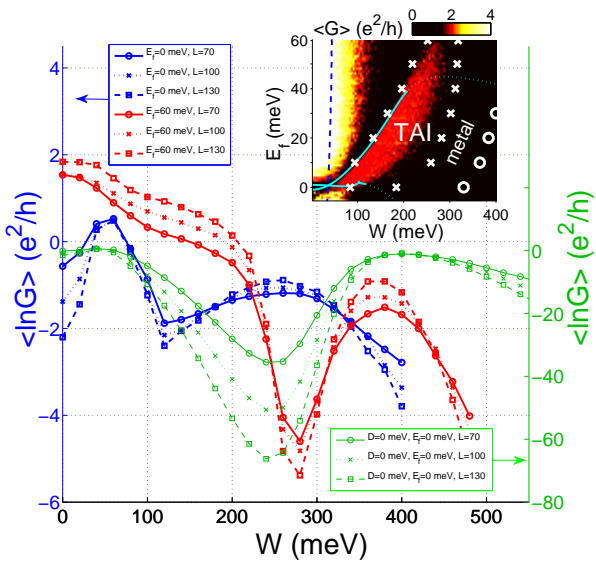}
\caption{\label{fig:epsart}(color online). Scaling behavior of
average logarithmic conductance as a function of $W$ at differnet
energies $E_f=60$ and 0 meV. The red and blue curves are calculated
with nonvanishing quadratic kinetic term while the green curves are
calculated by setting $D=0$. The blue and red curves are scaled by
the left axis, and the green curves are scaled by the right axis.
Inset: the Anderson transitions obtained from the scaling analysis
are marked by white symbols in the phase diagram.}
\label{AndersonTransition}
\end{figure}
To see the existence of such an Anderson transition, we carry out
the scaling analysis of the bulk conductance using periodic boundary
conditions \cite{MacKinnon}. The average logarithmic conductance,
$\langle \rm{Ln}G\rangle$, on $E_f-W$ plane is calculated for
different system sizes $L_x=L_y=L$, and the average is performed
under $2\times 10^3$ disorder realizations. For each $E_f$, the
Anderson transitions are recognized as the crossings of the $\langle
\rm{Ln}G(W)\rangle$ lines with different $L$'s, as seen in
Fig.\ref{AndersonTransition}. And in each region between two
crossings, the system is judged to be in insulating (metallic) if
$\langle \rm{Ln}G\rangle$ increases as the decrease (increase) of
$L$ at a fixed $W$. For example, there are three Anderson
transitions at $E_f=60$ meV as indicated by the red lines in
Fig.\ref{AndersonTransition}, where the system first transits from
metallic to insulating, which corresponds to the weak-disorder
boundary of the TAI phase obtained by SCBA; then transits from this
topological insulating phase to metallic again, which is the
localization-delocalization strong-disorder boundary of the TAI
phase we expect. After the third transition, the whole system is
totally localized and becomes an Anderson insulator. The complete
result is given in the inset of Fig.\ref{AndersonTransition}, where
the Anderson transitions are marked as white symbols on the phase
diagram. The region enclosed by the white crosses is the TAI island,
and we see that this region is much wider than the simulation result
for a finite-width system.

The conductance scaling at $E_f=0$ is also shown as blue lines in
Fig.\ref{AndersonTransition}. It is interesting to compare our
results with the recent work of Ref. \cite{Yamakage2010}, where only
a special energy point $E_f=0$ is considered, and the kinetic term,
$Dk^2$, is omitted. It is concluded there that in the presence of an
extra spin $s_z$-conservation breaking term, the Rashba spin-orbit
coupling, in the HgTe/CdTe QWs, a finite metallic region is found
which partitions the two topological distinct insulating phases.
However, our results demonstrate that the metallic region which
separates the TAI and the Anderson insulator phases can exist by
inclusion the quadratic kinetic term or tuning the $E_f$
to higher energies even with the spin $s_z$-conservation, and the
Rashba term is not necessary. This is the direct result of the
localization-delocalization Anderson transition we predict. In
Fig.\ref{AndersonTransition}, the scaling behavior of conductance at
$E_f=0$ by setting the parameter $D=0$ is shown as green lines,
where we see that the metallic region in-between the TAI and the
Anderson insulator phases disappear indeed, which is the case
discussed in Ref. \cite{Yamakage2010}.

\section {Conclusions}
In summary, the system of disordered HgTe/CdTe QWs is studied. The
evolution of its energy spectrum in the generalized momentum space
with the disorder strength is obtained, where the existence of
topologically protected helical edge states in the TAI phase is
demonstrated directly, which proves the conjecture that they are
responsible for the observed quantized conductance plateau. With
nonmagnetic perturbations which preserve the time-reversal symmetry
of the system, the edge states can be destroyed only through the
coupling with each other. For a finite-width system, it is shown
that the edge states decay in a {\it stepwise} pattern assisted with
the localized bulk states due to the fluctuations of the decay
length, which is in extraordinarily contrast with the exponential
decay of bulk states. While for an infinite system, the edge states
can become decoherent only through the coupling with extended
states, therefore a localization-delocalization Anderson transition
is expected in strong-disorder region, which is confirmed by the
scaling analysis. Moreover, multiple Anderson transitions are also
obtained, where in particular, a metallic region which separates two
topologically distinct insulating phases is found by inclusion of
the quadratic kinetic term or turning $E_f$ to higher energies
without an extra spin $s_z$-nonconservation term. These results are
also expected to be valid for 3D TAI \cite{Guo2010PRL}.

\acknowledgements
We would like to thank J. D. Zang, Z. Wang, C. X.
Liu for helpful discussions. This work is supported
by the NSFC under Grants No. 11004212, No. 10704080, No. 60877067,
and No. 60938004/F050802, by the STCSM under Project No.
08dj1400303, by the NBLXRYCY under Project No. 200901B3201015, and
by the NBNSFC under Project No. 2009A610060.

\end{document}